\documentclass[preprint, onecolumn, useamsfonts]{pasj01}
\Received{2021/07/29}
\Accepted{}
\draft
\onecolumn

\usepackage{natbib}
\usepackage{color}
\usepackage{bm}
\usepackage{subfigure}
\newcommand{\is}{i_{\star}}
\newcommand{\io}{i_{\rm orb}}
\newcommand{\vsini}{v_{\star}\sin{i_{\star}}}

\begin{document}
\title{
  Disentangling the stellar inclination
  of transiting planetary systems: fully analytic approach
  to the Rossiter-McLaughlin effect
  incorporating the stellar differential rotation}
\author{Shin \textsc{Sasaki} \altaffilmark{1}
and Yasushi \textsc{Suto} \altaffilmark{2,3}}
\email{sasaki@phys.se.tmu.ac.jp}
\altaffiltext{1}{Department of Physics, Tokyo Metropolitan University,
Hachioji, Tokyo 192-0397, Japan}
\altaffiltext{2}{Department of Physics, The University of Tokyo, Tokyo
113-0033, Japan}
\altaffiltext{3}{Research Center for the Early Universe, School of
Science, The University of Tokyo, Tokyo 113-0033, Japan}

\maketitle

\begin{abstract}
The Rossiter-McLaughlin (RM) effect has been widely used to estimate
the {\it sky-projected} spin-orbit angle, $\lambda$, of transiting
planetary systems.  Most of the previous analysis assume that the host
stars are rigid rotators in which the amplitude of the RM velocity
anomaly is proportional to $\vsini$.  When their latitudinal
differential rotation is taken into account, one can break the
degeneracy, and determine separately the equatorial rotation velocity
$v_\star$ and the inclination $\is$ of the host star.  We derive a
fully analytic approximate formula for the RM effect adopting a
parameterized model for the stellar differential rotation. For those
stars that exhibit the differential rotation similar to that of the
Sun, the corresponding RM velocity modulation amounts to several m/s.
We conclude that the latitudinal differential rotation offers a method
to estimate $\is$, and thus the full spin-orbit angle $\psi$, from the
RM data analysis alone.
\end{abstract}

%

\section{Introduction}

The spin-orbit angle $\psi$ of planetary systems is a unique probe of
their origin and evolution.  Observationally, $\psi$ can be estimated
for transiting planetary systems from the inclination angle of the
planetary orbit, $\io$, the projected spin-orbit angle, $\lambda$, and
the stellar inclination, $\is$, as
\begin{equation}
\label{eq:cospsi}
\cos \psi = \sin\is \sin\io \cos\lambda + \cos\is \cos\io .
\end{equation}
For the transiting planets, $\io \approx \pi/2$, and equation
(\ref{eq:cospsi}) is approximately given as
\begin{equation}
\label{eq:cospsi2}
\cos \psi \approx \sin\is \cos\lambda .
\end{equation}
Thus either $\sin\is \ll 1$ or $\cos\lambda \ll 1$ is a necessary
condition for the significant spin-orbit misalignment $\cos\psi \ll
1$.

For isolated and stable systems like our Solar system, $\psi$ is
supposed to keep the initial value at the formation epoch.  For
instance, in the simplest scenario, the star and the surrounding
protoplanetary disk are likely to share the same direction of the
angular momentum of the progenitor molecular cloud
\citep{Takaishi2020}. Therefore planets formed in the disk are
expected to have $\psi \sim 0$, as is the case for all the eight solar
planets ($\psi < 6^\circ)$.

Such a naive expectation, however, is not necessarily the case for the
observed distribution of $\lambda$ \citep{xue2014,winn2015}; more than
20 percents of transiting close-in gas-giant planets are misaligned,
in the sense that the 2$\sigma$ lower limit of their measured
$\lambda$ exceeds $30^\circ$ \citep{Kamiaka2019}.  While the origin of
such a misalignment is not yet well understood, possible scenarios
include planetary migration \citep{Lin1996}, planet-planet scattering
\citep{Rasio1996,Nagasawa2008,Nagasawa2011}, and perturbation due to
distant outer objects
\citep[e.g.,][]{Kozai1962,Lidov1962,Fabrycky2007,Batygin2012,xue2014,
  xue2017}.

We would like to stress that spin-orbit misaligned systems are often
defined by the value of $\lambda$ alone that are measured from the
Rossiter-McLaughlin (RM) effect, which is the radial velocity
  anomaly of the host star due to a distortion of the stellar line
  profiles during the planetary transit \citep[e.g.,][]{Rossiter1924,
    McLaughlin1924,Ohta2005,Winn2005}.

On the other hand, the stellar inclination may be inferred
either from asteroseismology \citep[e.g.,][]{Gizon2003} or from the
combination of the estimated stellar radius $R_\star$,
spectroscopically measured $(\vsini)$, and photometric stellar
rotation period $P_\star$:
\begin{equation}
\label{eq:is-combined}
\sin\is = \frac{(\vsini) P_\star}{2\pi R_\star}.
\end{equation}
The above two methods, however, do not always agree with each other,
and the reliable estimate has been made only to a limited number of
systems \citep{Kamiaka2018,Kamiaka2019}.

While $\lambda$ has been measured for more than 150 transiting
systems\footnote{https://www.astro.keele.ac.uk/jkt/tepcat/obliquity.html},
$\is$ are available from asteroseismology only for a couple of
systems, Kepler-25c and HAT-P-7 \citep{Benomar2014b}, and about thirty
systems have $\is$ estimated spectroscopically from equation
(\ref{eq:is-combined}); see \cite{Hirano2012} for instance. More
recently, \cite{Louden2021} derived a statistical distribution of
$\langle\sin\is\rangle$ for a sample of 150 Kepler stars with
transiting planets smaller than Neptune. Therefore, a complementary
method to estimate $\is$ breaking the degeneracy in $\vsini$ is
essential to understand the distribution of the spin-orbit angle
$\psi$ \citep[c.f.,][]{Albrecht2021}.

The RM effect of transiting planets was measured by \cite{Queloz2000}
for the first time. \cite{Ohta2005} derived an analytic approximate
formula for the RM effect, which were improved later by
\cite{Hirano2010} and \cite{Hirano2011}.  Such analytic expressions
for the RM effect are useful in understanding the parameter dependence
and degeneracy that are not easy to extract from numerical analysis.
Furthermore, \cite{Hirano2011} and \cite{Hirano2011b} attempted to
examine the effect of the differential rotation for the rapidly
rotating XO-3 system ($\vsini \approx 18$m/s and $\lambda \approx
40^\circ)$, but their result was not so constraining mainly due to the
large stellar jitter for such host stars ($\sim 15$m/s).

In the present paper, we improved the previous model for the RM effect
incorporating the differential rotation of the host star. Since our
model is fully analytic, one clearly understand how $\is$ and
$\lambda$ can be inferred from the RM modulation due to the
differential rotation; see equation (\ref{eq:Vz-diff-1}) below.  We
find that the differential rotation similar to the Sun produces an
additional feature of an amplitude on the order of several m/s in the
RM effect, which should be detectable for systems with good
signal-to-noise ratios.  Thus our analytic formula is useful in
breaking the degeneracy of $\vsini$, and allows the determination of
$\is$, and thus the true spin-orbit angle $\psi$, for transiting
planetary systems with the differential rotation of the host stars.

\section{Coordinates of the star and the transiting planet
  in the observer's frame}

\begin{figure}
 \begin{center}
 \includegraphics[width=10cm, bb = 0 0 732 386]{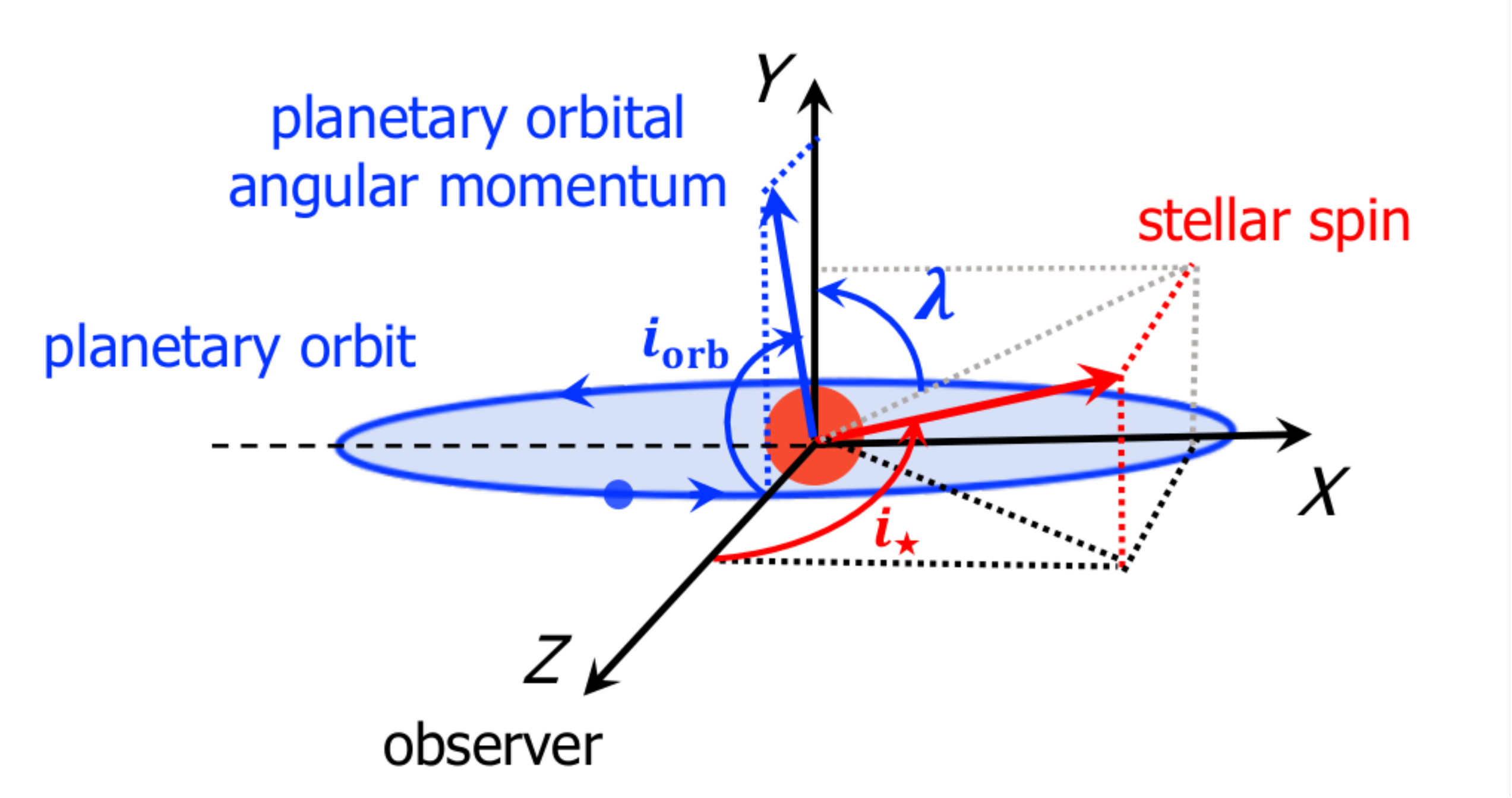} 
 \end{center}
 \caption{Spin-orbit architecture of transiting planetary systems in
   the observer's frame (O-frame); $\is$ and $\io$ are the stellar
   inclination and the planetary orbital inclination with respect to
   the observer, while $\lambda$ is the sky-projected spin-orbit angle.}
 \label{fig:osframes}
\end{figure}

We introduce three coordinate systems for the present analysis, whose
origin is chosen as the center of the host star.  The
observer's frame (O-frame) is represented by $(X, Y, Z)$.  The Z-axis
is directed toward the observer, and the Y-axis is chosen in such a
way that the orbital angular momentum vector of the planet lies on the
$YZ$ plane (Figure \ref{fig:osframes}).

\begin{figure}
 \begin{center}
  \includegraphics[width=10cm, bb = 0 0 753 519]{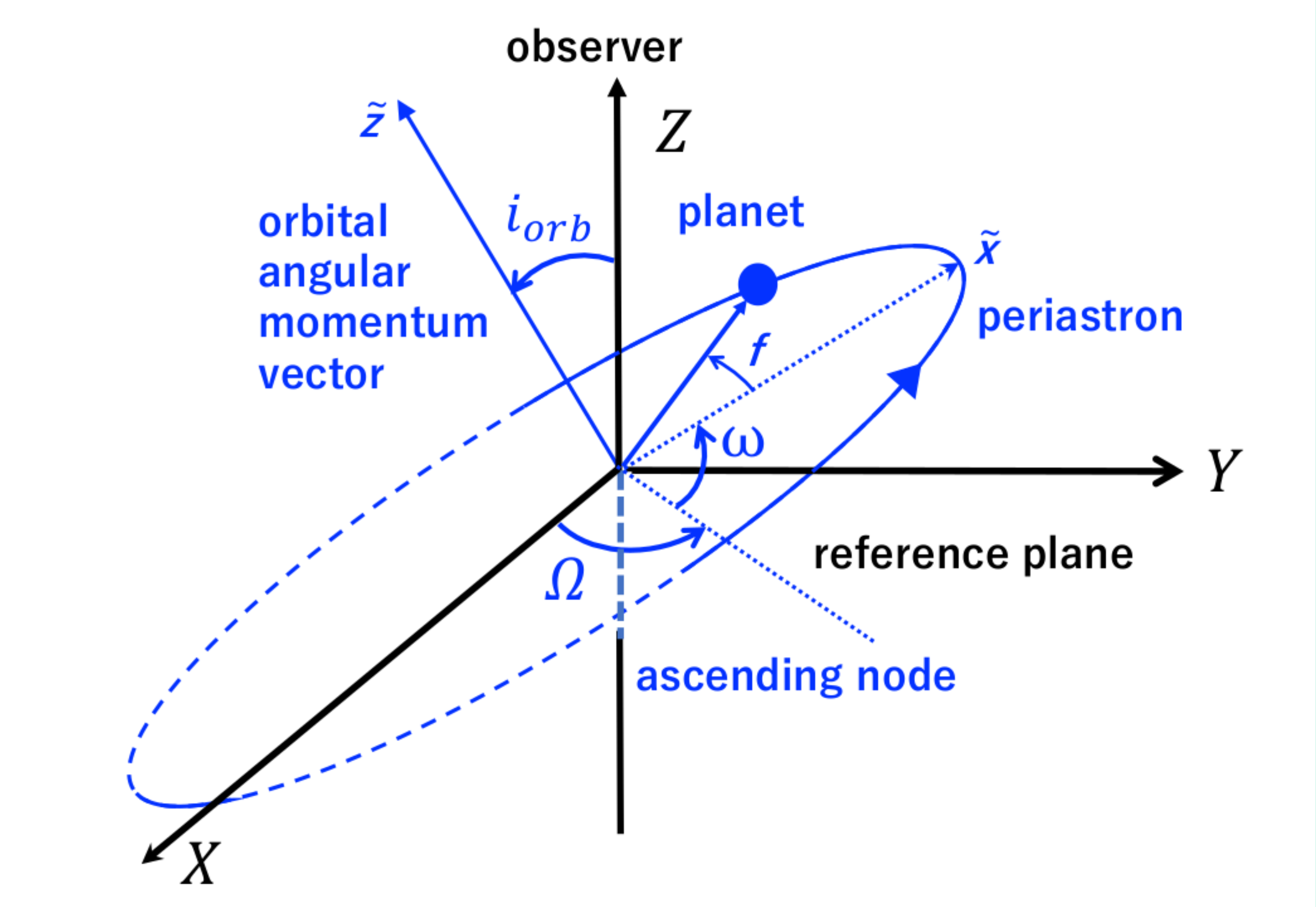} 
 \end{center}
 \caption{Planetary orbit in the observer's frame $(X,Y,Z)$ and
     in the planetary frame $(\tilde x, \tilde y, \tilde z)$.}
 \label{fig:opframes}
\end{figure}

The planetary frame (P-frame) is represented by $(\tilde{x},
\tilde{y}, \tilde{z})$. The planetary orbit is on the
$\tilde{x}\tilde{y}$ plane, thus the $\tilde{z}$-axis is along the
direction of the orbital angular momentum vector of the planet (Figure
\ref{fig:opframes}).  Finally, the stellar frame (S-frame) is
represented by $(x, y, z)$, and the $z$-axis is chosen as the
direction of the stellar spin vector.

An arbitrary vector $\bm{A}$ defined in the O-frame is
related  to $\bm{\tilde{a}}$ in the P-frame as
\begin{equation}
\label{eq:P2O}
  \left(
  \begin{array}{c} A_1 \\ A_2 \\ A_3 \end{array} \right)
  = R_3(\Omega) R_1(i_{\rm orb}) R_3(\omega)
  \left( \begin{array}{c} \tilde{a}_1 \\ \tilde{a}_2 \\ \tilde{a}_3
  \end{array}
 \right),
\end{equation}
where $R_i(\theta)$ denote the rotation matrix by angle $\theta$
around the $i$-th axis, and $i_{\rm orb}$, $\Omega$, $\omega$ are the
orbital inclination, the longitude of the ascending node, and the
argument of pericenter of the planet, respectively (Figure
\ref{fig:opframes}).

Similarly, $\bm{A}$ in the O-frame is related to $\bm{a}$ in the
  S-frame:
\begin{equation}
\label{eq:S2O}
  \left( \begin{array}{c} A_1 \\ A_2 \\ A_3 \end{array} \right)
  = R_3(-\lambda) R_1(-\is) R_3(\phi)
\left( \begin{array}{c} a_1 \\ a_2 \\ a_3 \end{array}  \right),
\end{equation}
where $\is$ and $\lambda$ denote the stellar inclination and the
projected spin-orbit angle, and $\phi$ is the azimuthal angle.  We
consider axi-symmetric stars in this paper, thus hereafter we fix
$\phi=0$ without loss of generality.

\section{Radial component of the stellar surface rotation velocity
  at the projected position of the transiting planet}

The position vector of a planet in the P-frame is expressed as
\begin{eqnarray}
  \label{eq:tilde-rp}
  \left(
  \begin{array}{c}
    \tilde{x}_p \\ \tilde{y}_p \\ \tilde{z}_p
  \end{array} \right)
  = \frac{a (1-e^2)}{1 + e \cos f}
\left( \begin{array}{c}
 \cos f \\
 \sin f \\ 0 \end{array} \right)
,
\end{eqnarray}
where $a$ and $e$ are the semi-major axis and eccentricity of the
planetary orbit, and $f$ is the true anomaly. Equation
(\ref{eq:tilde-rp}) is transformed to that in the O-frame using
equation (\ref{eq:P2O}):
\begin{equation}
\label{eq:Rp}
\left(
\begin{array}{c} X_p \\ Y_p \\ Z_p
\end{array} \right) = 
\frac{a(1-e^2) }{1+e \cos f}
\left(
\begin{array}{c}
  - \sin{\Omega } \sin{\left(f + \omega \right)} \cos{i_{\rm orb}}
  + \cos{\Omega} \cos{\left(f + \omega \right)}\\
  \sin{\Omega} \cos{\left(f + \omega \right)}
  + \sin{\left(f + \omega \right)} \cos{i_{\rm orb} } \cos{\Omega}\\
 \sin{i_{\rm orb}} \sin{\left(f + \omega \right)}
\end{array} \right)
.
\end{equation}

Essentially, the RM effect measures the line-of-sight component of the
stellar surface velocity at the projected surface position $(x_{sp},
y_{sp},z_{sp})$ in the S-frame that corresponds to $(X_p, Y_p, Z_p)$ in the
O-frame as the planet transits over the stellar surface. So we first
project $(X_p, Y_p, Z_p)$ in the O-frame on the position on the
stellar surface:
\begin{equation}
\label{eq:Rsp}
\left(
\begin{array}{c} X_{sp} \\ Y_{sp} \\ Z_{sp}
\end{array} \right) = 
\left(
\begin{array}{c}
  X_p\\
  Y_p\\
  \sqrt{R_\star^2-X_p^2-Y_p^2}
\end{array} \right)
,
\end{equation}
where
\begin{equation}
  Z_{sp}= \frac{a(1-e^2)}{1+e \cos f}
\sqrt{\frac{R_\star^2(1+e \cos f)^2}{a^2(1-e^2)^2}-
  \left[\sin^2(f + \omega) \cos^2\io  + \cos^2(f + \omega) \right]} .
\end{equation}
Next, we rotate ($X_{sp}$, $Y_{sp}$, $Z_{sp}$) to that in the S-frame
using equation (\ref{eq:S2O}):
\begin{eqnarray}
  \label{eq:P2O2S}
\left(
\begin{array}{c}
  x_{sp} \\ y_{sp} \\ z_{sp}
\end{array}
\right)
  &=& R_1(+\is)R_3(+\lambda) 
\left(
\begin{array}{c}
  X_{sp} \\ Y_{sp} \\ Z_{sp}
\end{array}
\right) \cr
&=&
\left(
\begin{array}{ccc}
  1 & 0 & 0\\
  0 & \cos\is & -\sin\is\\
  0 & \sin\is & \cos\is \\
\end{array}
\right) 
\left(
\begin{array}{ccc}
  \cos\lambda & -\sin\lambda & 0\\
  \sin\lambda & \cos\lambda & 0\\
  0 & 0 & 1\\
\end{array}
\right) 
\left(
\begin{array}{c}
  X_{p} \\ Y_{p} \\ Z_{sp}
\end{array}
\right) .
\end{eqnarray}
Specifically, we obtain the planetary position in the S-frame in terms
of the orbital elements:
\begin{eqnarray}
\label{eq:xp}
x_{sp} &=&
 \frac{a(1-e^2)}{1+e\cos f}
\Big[\cos(\lambda+\Omega)\cos (f+\omega)-
  \cos\io \sin(\lambda+\Omega)\sin (f+\omega)\Big], \\
y_{sp} &=&
 \frac{a(1-e^2)}{1+e\cos f}
    \Big[\cos\is \left(\sin(\lambda+\Omega)\cos(f+\omega)
       +\cos(\lambda+\Omega)\cos\io\sin(f+\omega)\right) \cr 
      && \qquad\qquad\qquad  -\sin\is~Z_{sp} \Big], \\
\label{eq:zp}
z_{sp} &=& \frac{a(1-e^2)}{1+e\cos f}
\sin\is\left[\sin(\lambda+\Omega)\cos(f+\omega)
  +\cos(\lambda+\Omega)\cos\io\sin(f+\omega)\right]\cr
       && \qquad\qquad\qquad  +\cos\is~ Z_{sp} .
\end{eqnarray}

Consider a point on the stellar surface at latitude $\ell$ and longitude
$\varphi$ in the S-frame:
\begin{eqnarray}
  \label{eq:rs}
  \left( \begin{array}{c} x \\ y \\ z \end{array} \right)
  = R_\star
\left( \begin{array}{c}
 \cos \ell \cos\varphi \\
 \cos \ell \sin\varphi\\
 \sin \ell \end{array} \right)
.
\end{eqnarray}
If the stellar surface rotates around the $z$-axis, the corresponding
velocity in the S-frame is given by
\begin{equation}
\label{eq:vs}
\left( \begin{array}{c} v_x \\ v_y \\ v_z \end{array} \right) =
R_\star \dot{\varphi}(\ell)
\left( \begin{array}{c} - \cos l \sin \varphi \\
\cos l \cos \varphi \\ 0 \end{array} \right)
= \dot{\varphi}(\ell)
\left( \begin{array}{c} - y \\ x \\ 0 \end{array} \right)
.
\end{equation}

In what follows, we adopt the following parameterized model for the
latitudinal differential rotation:
\begin{equation}
\label{eq:wl}
  \dot{\varphi} (\ell) = \omega_0 (1 - \alpha_2 \sin^2\ell -
\alpha_4 \sin^4\ell).
\end{equation}
For the Sun, $\omega_{0\odot}\approx 2.972\times
10^{-6}\mathrm{~rad~s^{-1}}$, $\alpha_{2\odot} \approx 0.163$, and
$\alpha_{4\odot} \approx 0.121$ \citep{1990ApJ...351..309S}.  
  The differential rotation of stars is not so accurately measured
  except for the Sun.  \citet{Brun2017} presented a model from their
  numerical simulations in which the amplitude of the latitudinal
  differential rotation changes by about 20 percent, roughly
  consistent with the Sun. It is not clear, however, to what extent
  the result is applicable for other stars.

The stellar surface velocity (\ref{eq:vs}) in the S-frame is
transformed into that in the O-frame using equation (\ref{eq:S2O}):
\begin{equation}
\label{eq:VO}
  \left( \begin{array}{c} V_X \\ V_Y \\ V_Z \end{array} \right)
    = R_3(-\lambda) R_1(-\is) 
    \left( \begin{array}{c} v_x \\ v_y \\ v_z \end{array} \right)
    = \dot{\varphi}(\ell) \left( \begin{array}{c}
    -y \cos \lambda + x \sin \lambda \cos \is \\
    y \sin \lambda + x \cos \is \cos \lambda \\
    -x \sin \is
    \end{array} \right) .
\end{equation}

Finally, equations (\ref{eq:xp}) and (\ref{eq:VO}) yield
\begin{eqnarray}
\label{eq:Vz-rp}
  && V_{Z}(\bm{r}_{sp}) = -x_{sp} \dot{\varphi}(\ell) \sin\is \cr
  &=& \frac{a(1-e^{2})}{1+e\cos f}
  \dot{\varphi}(\ell)\sin\is
      [\cos\io \sin(\lambda+\Omega) \sin(f + \omega)
 - \cos(\lambda+\Omega)\cos(f + \omega)] ,
\end{eqnarray}
where the latitude $\ell$ is computed from equations (\ref{eq:zp})
and (\ref{eq:rs})
\begin{eqnarray}
\label{eq:sinl-1}
  && \frac{R_\star(1+e\cos f)}{a(1-e^2)} \sin\ell
  =  \sin\is\left(\sin(\lambda+\Omega)\cos(f+\omega)
       +\cos(\lambda+\Omega)\cos\io\sin(f+\omega)\right) \cr
&& \qquad\qquad\qquad  +\cos\is
\sqrt{\frac{R_\star^2(1+e \cos f)^2}{a^2(1-e^2)^2}-
  \left[\cos^2\io \sin^2(f + \omega)  + \cos^2(f + \omega) \right]} 
.
\end{eqnarray}

\section{Differential rotation effect for a circular planetary orbit}

Equation (\ref{eq:Vz-rp}) with equation (\ref{eq:sinl-1}) provides the
key result that disentangles $\is$ through the differential rotation
of the host stars using the RM effect alone.  For definiteness, we focus
on a circular orbit ($e=0$, $\omega=0$ and $\Omega=\pi$) in this
section.  Then, equation (\ref{eq:sinl-1}) is simplified as
\begin{eqnarray}
\label{eq:sinl-2}
\frac{R_\star}{a} \sin\ell
 &=& \cos\is
\sqrt{\frac{R_\star^2}{a^2}-
  \left(\cos^2\io \sin^2 f  + \cos^2 f \right)}\cr
&& \qquad\qquad\qquad - \sin\is\left(\sin\lambda\cos f
       +\cos\lambda\cos\io\sin f\right) .
\end{eqnarray}

Since $(X_p, Y_p) \approx (-a\cos f, -a\cos\io)$, the true anomaly
during the transit ($X_p^2+Y_p^2<R_\star^2$) may be approximated as
\begin{eqnarray}
  f \equiv  \frac{\pi}{2} + \Delta f,
\end{eqnarray}
where
\begin{eqnarray}
-\sqrt{\left(\frac{R_\star}{a}\right)^2-\cos^2\io}
  <\Delta f <\sqrt{\left(\frac{R_\star}{a}\right)^2-\cos^2\io}.
\end{eqnarray}
We define the dimensionless impact parameter $b$ and dimensionless
time variable $\tau$:
\begin{eqnarray}
  \cos\io &\equiv& \frac{R_\star}{a} b \qquad\quad (0<b<1), \\
  \Delta f &\equiv& \frac{R_\star \sqrt{1-b^2}}{a} \tau \quad (-1<\tau<1).
\end{eqnarray}
Thus equation (\ref{eq:sinl-2}) reduces to
\begin{eqnarray}
\label{eq:sinl-3}
\sin\ell
 &\approx& \cos\is \sqrt{1-b^2}\sqrt{1-\tau^2}
 - \sin\is\left(\tau \sqrt{1-b^2} \sin\lambda
       +b \cos\lambda\right) .
\end{eqnarray}
Equation (\ref{eq:Vz-rp}) with equation (\ref{eq:wl}) reduces to
\begin{eqnarray}
\label{eq:Vz-circular}
V_Z(\tau)
&\approx&
R_\star \omega_0\sin\is (1 - \alpha_2\sin^2 \ell - \alpha_4\sin^4 \ell) 
\left(\tau\sqrt{1-b^2}\cos\lambda + b \sin\lambda\right) \cr
&\equiv& V_{\rm rigid}(\tau)+V_{\rm diff}(\tau),
\end{eqnarray}
where the rigid rotation component is
\begin{eqnarray}
\label{eq:Vz-rigid}
V_{\rm rigid}(\tau) \equiv \vsini
\left(\tau\sqrt{1-b^2}\cos\lambda + b \sin\lambda\right),
\end{eqnarray}
and the additional modulation term due to the differential rotation is
\begin{eqnarray}
\label{eq:Vz-diff}
V_{\rm diff}(\tau) \equiv
&& -(\alpha_2 \sin^2\ell+\alpha_4 \sin^4\ell) V_{\rm rigid} \cr
= && - \Big\{
  \alpha_2  \Big[\cos\is \sqrt{1-b^2}\sqrt{1-\tau^2}
 - \sin\is\left(\tau \sqrt{1-b^2} \sin\lambda
       +b \cos\lambda\right)\Big]^2 \cr
&& + \alpha_4  \Big[\cos\is \sqrt{1-b^2}\sqrt{1-\tau^2}
 - \sin\is\left(\tau \sqrt{1-b^2} \sin\lambda
       +b \cos\lambda\right)\Big]^4 \Big\} \cr
  && \qquad\qquad
\times \vsini\left(\tau\sqrt{1-b^2}\cos\lambda + b \sin\lambda\right).
\end{eqnarray}

If we assume a rigid rotation for the central star, time-variation of
equation (\ref{eq:Vz-rigid}) is specified by $\vsini$ and $\lambda$
(the impact parameter $b$ is estimated from the transit
lightcurve). The differential rotation term, equation
(\ref{eq:Vz-diff}), can break the degeneracy of $\vsini$, and enables
us to estimate $\alpha_2$, $\alpha_4$ and $\is$ separately in
principle, through, for instance, the measurements at the central
transit ($\tau=0$) and egress/ingress ($\tau=\pm1$) phases:
\begin{eqnarray}
\label{eq:Vz-diff-0}
\frac{V_{\rm diff}(0)}{\vsini} 
&=& - b\sin\lambda
(b \sin\is\cos\lambda -\sqrt{1-b^2}\cos\is)^2\cr
&& \qquad \times [\alpha_2 + \alpha_4
(b \sin\is\cos\lambda -\sqrt{1-b^2}\cos\is)^2], \\
\label{eq:Vz-diff-1}
\frac{V_{\rm diff}(\pm1)}{\vsini}
&=& -\sin^2\is
(b \cos\lambda \pm \sqrt{1-b^2}\sin\lambda)^2
(b \sin\lambda \pm \sqrt{1-b^2}\cos\lambda) \cr
&& \qquad \times[\alpha_2 +\alpha_4 \sin^2\is
(b \cos\lambda \pm \sqrt{1-b^2}\sin\lambda)^2].
\end{eqnarray}
Strictly speaking, however, the RM effect does not measure directly
$V_Z(\bm{r}_{sp})$, but its convolution over the stellar disk including
the limb darkening, which will be taken into account in the next section.

\section{Analytic approximation to the Rossiter-McLaughlin effect
  for differentially rotating stars}

\subsection{The Rossiter-McLaughlin effect in Gaussian approximation}

The RM radial velocity anomaly $\Delta v_{\rm RM}$ is
roughly given by 
\begin{equation}
\label{eq:RM-OTS}
\Delta v_{\rm RM} = - \frac{f}{1-f} v_p,
\end{equation}
where $v_p$ is the line-of-sight velocity of the stellar surface
  element occulted by the planet, and $f (\ll 1)$ represents the
fraction of the occulted area in units of the stellar disk area
\citep{Ohta2005}. Equation (\ref{eq:RM-OTS}) does not take into
account the stellar line profile due to the thermal and rotational
broadenings, and micro/macro-turbulences.

\cite{Hirano2010} and \cite{Hirano2011} improved equation
(\ref{eq:RM-OTS}), and derived the following analytic formula under
the Gaussian line profile approximation:
\begin{eqnarray}
\label{eq:RM-Gaussian}
\Delta v_{\rm RM}
= - \left(\frac{2\beta_\star^2}{\beta_\star^2+\beta_p^2}\right)^{3/2}
f v_p \exp\left(-\frac{v_p^2}{\beta_\star^2+\beta_p^2}\right),
\end{eqnarray}
where
\begin{eqnarray}
\label{eq:beta-p}
  \beta_p^2 &=& \beta^2+\zeta^2, \\
\label{eq:beta-*}
  \beta_\star^2 &=& \beta^2+\zeta^2+\sigma_\star^2, \\
\label{eq:sigma-*}
  \sigma_\star^2 &\approx& (\vsini/1.3)^2 .
\end{eqnarray}
In the above expressions, $\beta$ represents the thermal broadening
and micro-turbulence due to the convection, and $\zeta$
corresponds to macroturbulence that may be approximated as
\begin{eqnarray}
  \zeta \approx \left( 3.98 +
  \frac{T_{\rm eff}-5770 {\rm K}}{650 {\rm K}}\right) ~{\rm km/s}
\end{eqnarray}
\citep{Valenti2005,Hirano2012}. 

The stellar lines are additionally broadened by the stellar rotation,
whose Gaussian width is found to be approximated as equation
(\ref{eq:sigma-*}) by \cite{Hirano2010}.  Note that \cite{Hirano2010}
and \cite{Hirano2011} adopted the Gaussian profile $\propto
e^{-x^2/\sigma^2}$ instead of $\propto e^{-x^2/2\sigma^2}$.  Thus
equations (\ref{eq:RM-Gaussian}) to (\ref{eq:sigma-*}) are slightly
different from equation (2) in \cite{Boue2013}, for instance.

\begin{figure}
 \begin{center}
  \includegraphics[width=10cm, bb = 0 0 560 414]{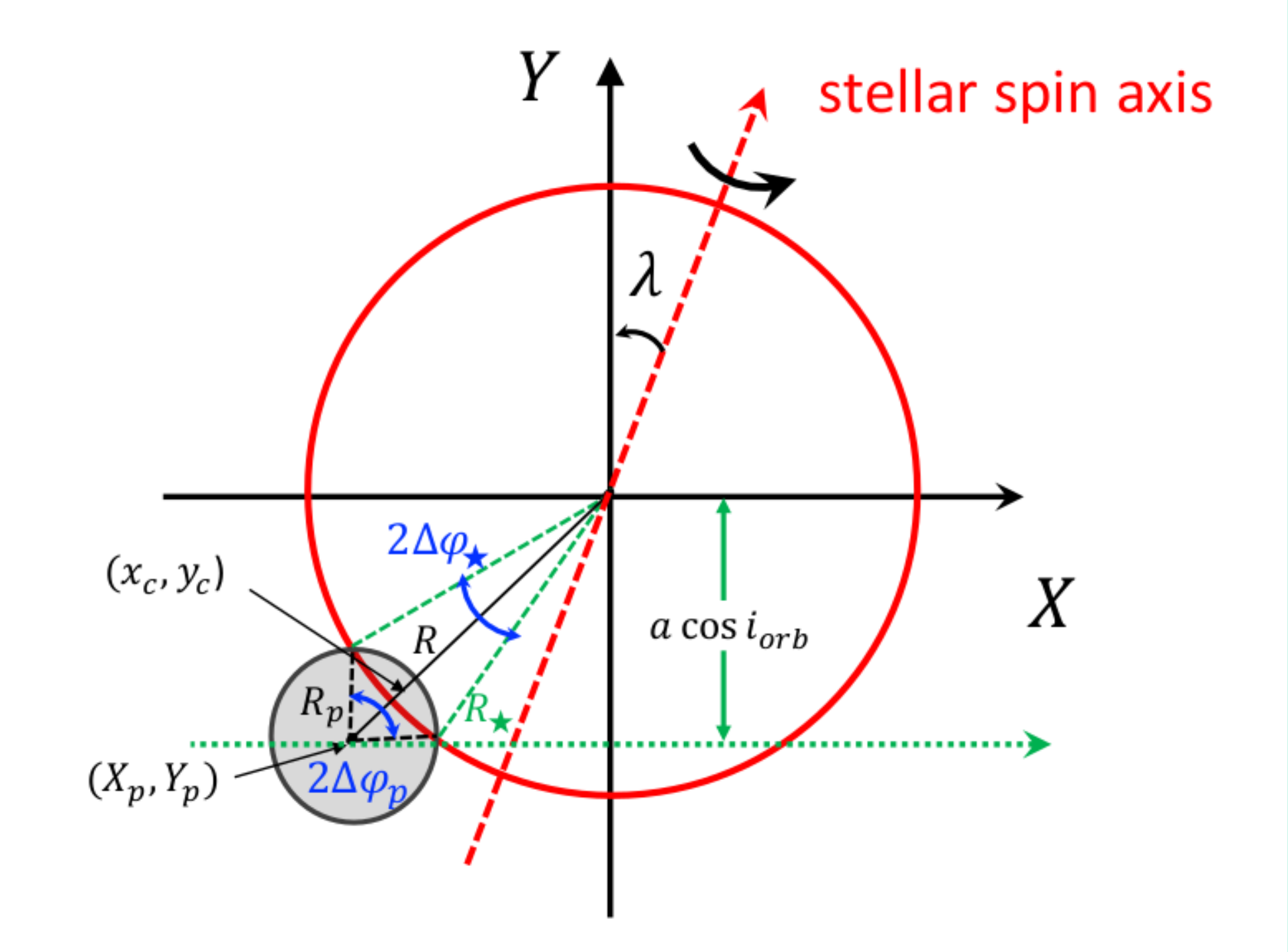} 
 \end{center}
 \caption{Schematic illustration of the stellar
   disk occulted by a transiting planet at the ingress/egress phases.}
 \label{fig:ingress-egress}
\end{figure}

The fraction of the occulted area neglecting the limb darkening can be
computed from equations (B1) to (B3) in \cite{Hirano2010}.  Figure
\ref{fig:ingress-egress} shows relevant angles in the O-frame in which
the centers of the star and planet are located at $(X,Y)=(0,0)$ and
$(X_p, Y_p)$ given by equation (\ref{eq:P2O2S}):
\begin{eqnarray}
  \cos\Delta\varphi_\star &=& \frac{R_\star^2+R^2-R_p^2}{2 R_\star R},\\
  \cos\Delta\varphi_p &=& \frac{R_p^2+R^2-R_\star^2}{2R_p R},\\
  R &=& \sqrt{X_p^2+Y_p^2},
\end{eqnarray}
with $R_p$ being the radius of the planet.
Then the fraction $f_0(R)$ neglecting limb darkening is given as
\begin{eqnarray}
  \hspace*{-1.5cm}
  f_0(R) = \left\{
  \begin{array}{ll}
   0 & (R>R_\star+R_p) \\
   \displaystyle
     \frac{1}{\pi}
   \left[\Delta\varphi_\star -\frac{\sin2\Delta\varphi_\star}{2}
     + \frac{R_p^2}{R_\star^2}
     \left(\Delta\varphi_p -\frac{\sin2\Delta\varphi_p}{2}\right)
     \right] & (R_\star-R_p \leq R \leq R_\star+R_p) \\
   \displaystyle \frac{R_p^2}{R_\star^2} & (R<R_\star-R_p) \\
  \end{array} \right.
\end{eqnarray}

\subsection{Effect of Limb-darkening}

We adopt the quadratic limb darkening law in the S-frame:
\begin{eqnarray}
f_{\rm LD}(x,y) &=& 
\frac{I(x, y)}{I(0,0)} =
1 - 2 q_2 \sqrt{q_1} (1 - \mu) - \sqrt{q_1} (1 - 2 q_2) (1-\mu)^2, \\
\mu &=& \sqrt{1 - \frac{x^2+y^2}{R_\star^2}} ,
\end{eqnarray}
where $I(x,y)$ is the stellar intensity at a location $(x,y)$
\citep{2013MNRAS.435.2152K}.  
The parameters $q_1$ and $q_2$ are related to $u_1, u_2$ which are used in
\cite{Hirano2010} by the following equation:
\begin{equation}
    q_1 = (u_1+u_2)^2, ~~ q_2 = \frac{u_1}{2 (u_1 + u_2)}.
\end{equation}

We approximate the effect of the limb darkening of the occulted region
by the value evaluated at $(x_c,y_c)$, instead of integrating over the
corresponding area:
\begin{eqnarray}
\left( \begin{array}{c} x_c \\ y_c \end{array} \right)
= \left\{
  \begin{array}{cc}
    \displaystyle
    \frac{R_\star+R-R_p}{2r}
    \left( \begin{array}{c} X_p \\ Y_p \end{array} \right)
    & \qquad\ (R_\star-R_p \leq R \leq R_\star+R_p) \\
    \left( \begin{array}{c} X_p \\ Y_p \end{array} \right)
 &  (R < R_\star-R_p) \\
    \end{array} \right.
\end{eqnarray}
Thus we use
\begin{eqnarray}
\label{eq:f-formula}
f(R) \approx f_{\rm LD}(x_c, y_c) f_0(X_p,Y_p)
\left[ 1 - \frac{2 \sqrt{q_1} q_2}{3}
    - \frac{\sqrt{q_1} (1 - 2 q_2)}{6} \right]^{-1}.
\end{eqnarray}
Also we adopt equation (\ref{eq:Vz-rp}) evaluated at $(x_c,y_c)$
for $v_p$ in equation (\ref{eq:RM-Gaussian}):
\begin{equation}
\label{eq:vp-formula}
v_p(\bm{r}_{sp}) = V_Z(x_c,y_c).
\end{equation}

\section{Results}

\begin{table}
  \tbl{Adopted parameters for the system considered in section 6.}{%
  \begin{tabular}{llll}
  Name & Symbol & Fiducial Value & reference \\ 
  \hline
  \hline
  star & & & \\
  \hline
  radius & $R_{\star}$ & $1.14 R_{\solar}$ & \mbox{\cite{2021A&A...647A..26C}} \\
  spin velocity & $v_{\star} \sin i_{\star} $ & 4.5 km/s & 
  \mbox{\cite{Hirano2011}} \\
  Gaussian fitting & $\beta$ & 4.0 km/s & 
  \mbox{\cite{Hirano2011}} \\
   & $\zeta$ & 4.3 km/s & \mbox{\cite{Hirano2011}} \\
   & $\sigma_{\star} $ & $ \frac{v_{\star} \sin i_{\star}}{1.3}$ &
   \mbox{\cite{Hirano2011}} \\
 limb darkening & $(q_1, q_2)$ & (0.56, 0.32) & 
 \mbox{\cite{Hirano2010}} \\
 differential rotation & $(\alpha_2, \alpha_4)$ & (0.163, 0.121) & \mbox{\cite{1990ApJ...351..309S}} Solar value \\
  \hline
  \hline
 planet & & & \\
  \hline
 radius ratio & $R_p/R_{\star}$ & 0.121 &  
 \mbox{\cite{2021A&A...647A..26C}} \\
 \hline
 \hline
 orbit & & & \\
 \hline
 orbital period & $P_{\rm orb} $ & 3.5 days & 
 \mbox{\cite{2021A&A...647A..26C}} \\
 semi-major axis & $a$ & 0.047 au & 
 \mbox{\cite{2021A&A...647A..26C}} \\
\end{tabular}}\label{tab:paramvalue}
\begin{tabnote}
We adopt fiducial values of parameters mainly from the HD209458 system
except for  $\io$, $\is$, $\lambda$, and the differential rotation
parameters ($\alpha_2$, $\alpha_4$).  For simplicity, we consider a
circular orbit ($e=0$), and thus fix $\omega=0$ and $\Omega=\pi$
without loss of generality.
\end{tabnote}
\end{table}

Equation (\ref{eq:RM-Gaussian}) with equations (\ref{eq:f-formula})
and (\ref{eq:vp-formula}) provides an analytic formula for the RM
velocity anomaly taking account of the differential rotation of the
host star.

In order to examine to what extent one can break the degeneracy of
$\is$ using the differential rotation, we assume a set of fiducial
values for system parameters (Table \ref{tab:paramvalue}).  They are
adopted from the HD209458 system, except for $\io$, $\is$, $\lambda$,
and the differential rotation parameters ($\alpha_2$, $\alpha_4$),
which we vary in the analysis below.  For simplicity, we consider a
circular orbit and fix $\omega=0$ and $\Omega=\pi$.  Since we fix the
equatorial stellar velocity $\vsini=4.5$km/s, the equatorial angular
velocity is assumed to vary as $\sin\is$ as
\begin{equation}
\omega_0=\frac{4.5~{\rm km/s}}{R_\star\sin\is}.
\end{equation}

\begin{figure}
 \begin{center}
  \includegraphics[width=12cm, bb = 0 0 1024 910]{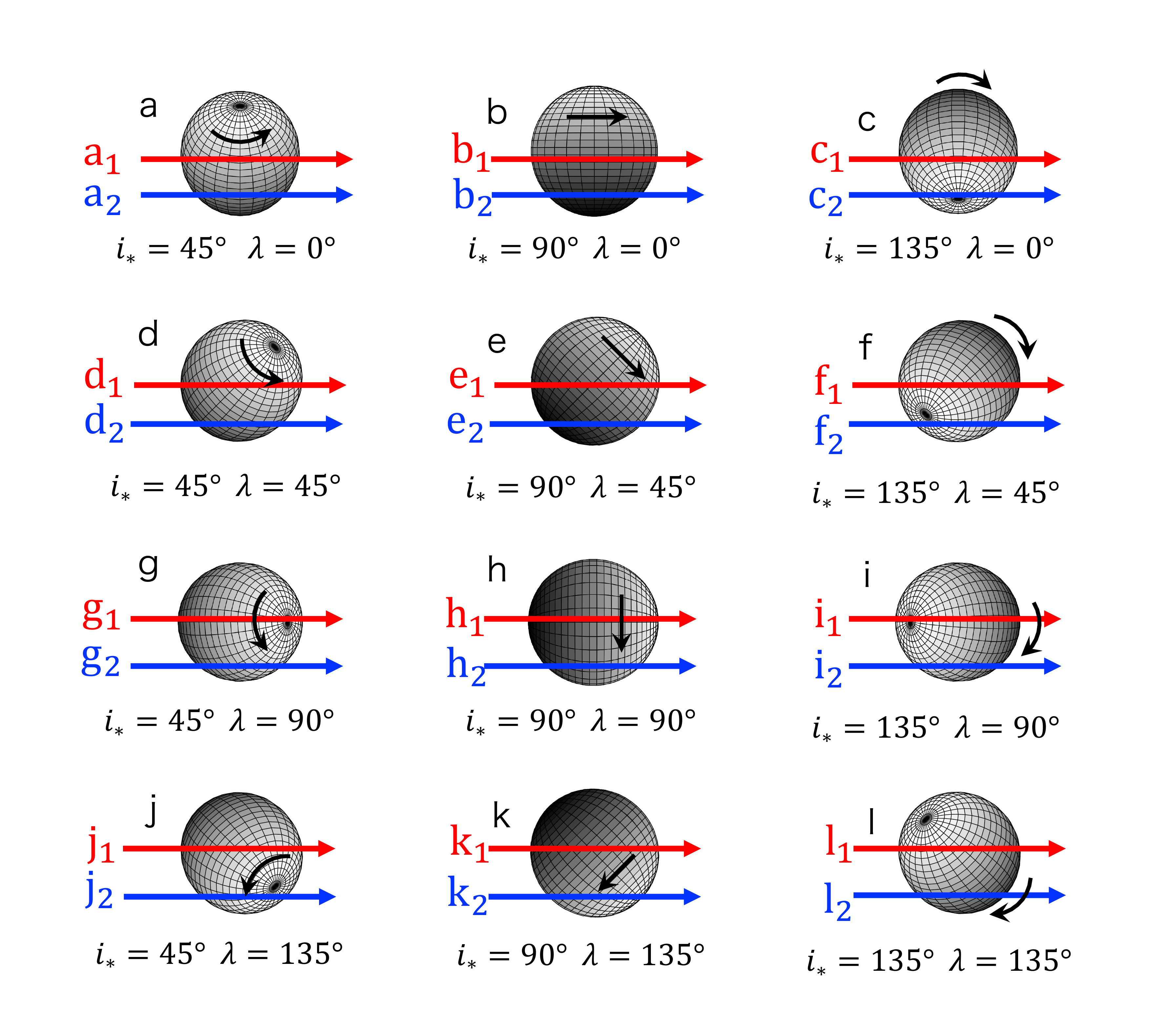} 
 \end{center}
 \caption{Schematic trajectories of transiting planets on the stellar
   disk for different $\is$ and $\lambda$. Upper and lower arrows in
   each panel illustrate trajectories for $\io=90^\circ (b=0)$ and
   $\io<90^\circ (b \not=0)$.}
 \label{fig:planet-trajectories}
\end{figure}

\subsection{Dependence on the differential rotation parameters}

\begin{figure}
 \begin{center}
  \includegraphics[width=12cm, bb = 0 0 1127 892]{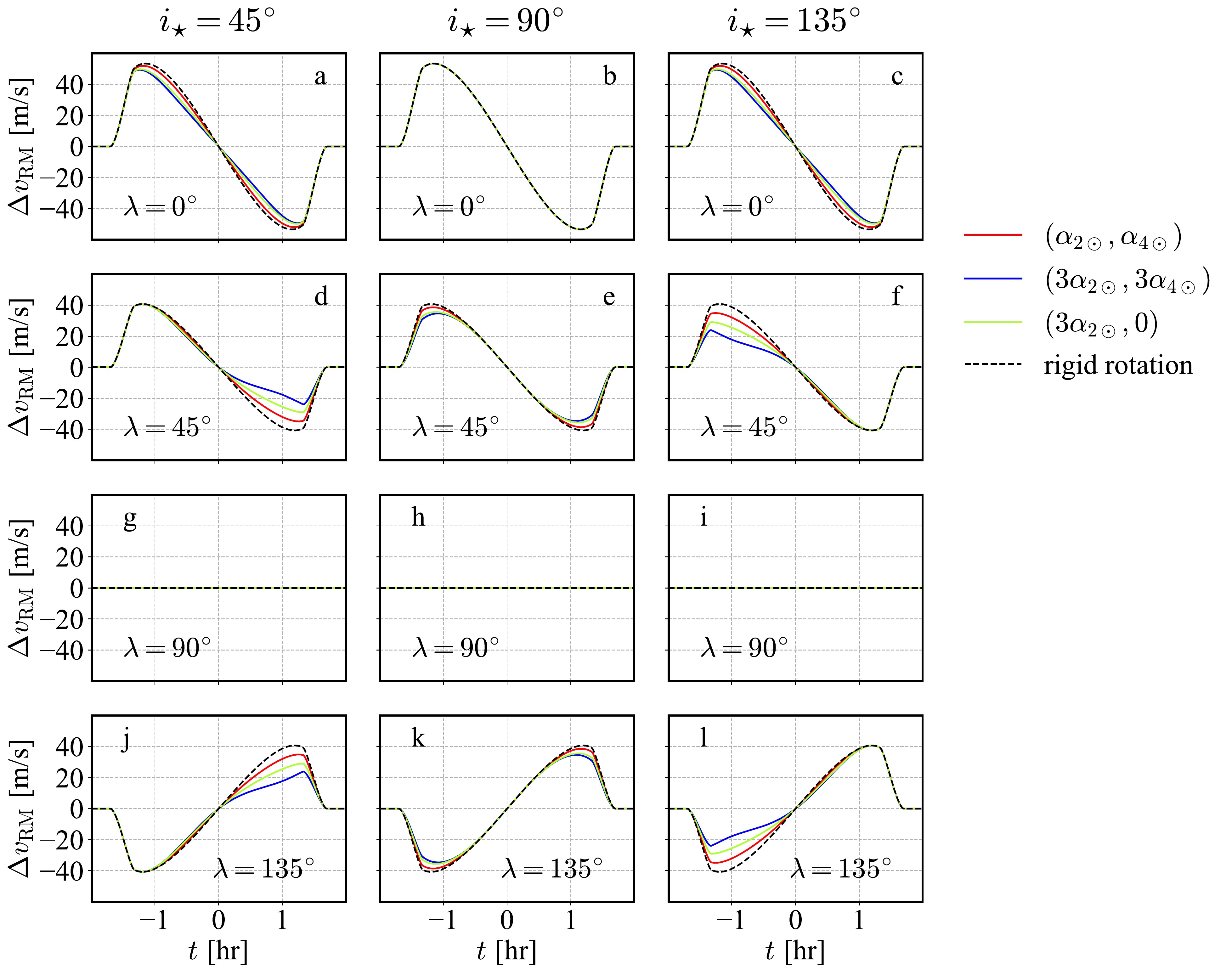} 
 \end{center}
 \caption{Effect of the stellar latitudinal differential rotation on
   the RM velocity anomaly $\Delta v_{\rm RM}(t)$ for $i_{\rm orb}
   =90^\circ (b=0)$ for different values of $\lambda$ and $\is$.  Each
   panel corresponds to the upper trajectories shown in Figure
   \ref{fig:planet-trajectories}.  Red, blue, and greenyellow curves
   are for $(\alpha_2,\alpha_4)=(\alpha_{2\odot},\alpha_{4\odot})$,
   $(3\alpha_{2\odot},3\alpha_{4\odot})$, and $(3\alpha_{2\odot},0)$,
   respectively. Dashed curves indicate the case for rigid rotation
   ($\alpha_2=\alpha_4=0$). We do not plot the case for
   $(\alpha_2,\alpha_4)=(0, 3\alpha_{4\odot})$ because it is very
   close to models with
   $(\alpha_2,\alpha_4)=(\alpha_{2\odot},\alpha_{4\odot})$.}
 \label{fig:I90}
\end{figure}

\begin{figure}
 \begin{center}
  \includegraphics[width=12cm, bb = 0 0 1127 892]{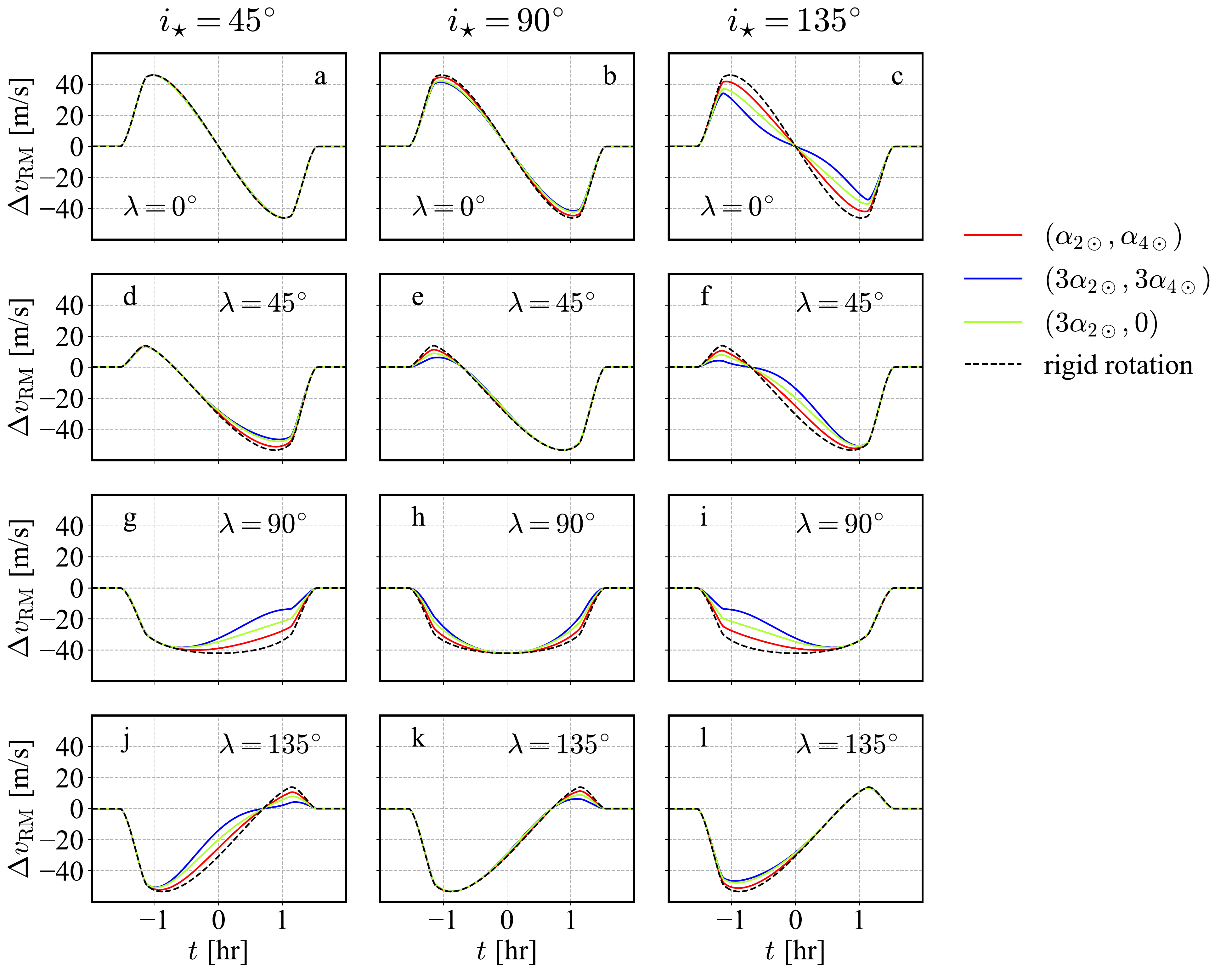} 
 \end{center}
 \caption{Same as Figure \ref{fig:I90} except 
   for $\io =87^\circ (b \approx 0.47)$.}
\label{fig:I87}
\end{figure}

Figure \ref{fig:planet-trajectories} illustrates schematic
trajectories of planets during their transits for different
combinations of the spin-orbit angles ($\is$,$\lambda$).  The
corresponding RM velocity anomaly $\Delta v_{\rm RM}$ for each panel
is plotted in Figures \ref{fig:I90} and \ref{fig:I87} for
$\io=90^\circ (b=0)$ and $87^\circ (b\approx0.47)$, respectively,
for different combinations of $\alpha_2$ and $\alpha_4$.

First, let us consider panels a to c in Figure \ref{fig:I90}, in which
$i_{\rm orb} = 90^{\circ}$ and $\lambda=0$. Since planets in panels a
and c move away from the stellar equator (Figure
\ref{fig:planet-trajectories}), the corresponding $\Delta v_{\rm
  RM}(t)$ with differential rotation becomes smaller than that for
rigid rotator. Due to the symmetry between $l \leftrightarrow -l$ in
equation (\ref{eq:wl}), $\Delta v_{\rm RM}(t)$ in panels a and c is
identical. In contrast, panel b corresponds to the planetary
trajectory along the equator (Figure \ref{fig:planet-trajectories}),
$\Delta v_{\rm RM}(t)$ is not affected by the presence of differential
rotation.

Similarly, one may understand the behavior of panels d to f
($\lambda=45^\circ$) by comparing with the corresponding trajectories
in Figure \ref{fig:planet-trajectories}. Since the trajectory $d_1$
moves from low to high latitude regions, the effect of differential
rotation is stronger after the central transit epoch as shown in
Figure \ref{fig:I90}. The trajectory $f_1$ is opposite, and the
trajectory $e_1$ corresponds to the case between the two cases.

Panels j to l ($\lambda=135^\circ$) are easily understood from panels d
to f since equations (\ref{eq:Vz-rp}) and (\ref{eq:sinl-1})
indicates $V_z(\lambda)=-V_z(\pi - \lambda)$ for $\io=90^\circ$. Finally,
panels g to i ($\lambda=90^\circ$) correspond to the trajectories with
$V_z=0$ for $\io=90^\circ$.

Consider next the case for $\io=87^\circ (b \approx 0.47)$, {\it i.e.}, Figure
\ref{fig:I87} and lower trajectories in the corresponding panels of
Figure \ref{fig:planet-trajectories}. In this case, the symmetry
between $\lambda \leftrightarrow \pi - \lambda$ no longer holds, and
$\Delta v_{\rm RM}(t)$ exhibits small but clear dependence on $\io$,
$\is$, and $\lambda$.

In conclusion, Figures \ref{fig:I90} and \ref{fig:I87} imply that
the differential rotation can be used to estimate $\is$ from the
RM velocity anomaly curve through the latitudinal dependence of
$\dot\varphi(\ell)$ in equations (\ref{eq:Vz-rp}) and
(\ref{eq:sinl-1}).

\begin{figure}
 \begin{center}
  \includegraphics[width=12cm, bb = 0 0 1109 892]{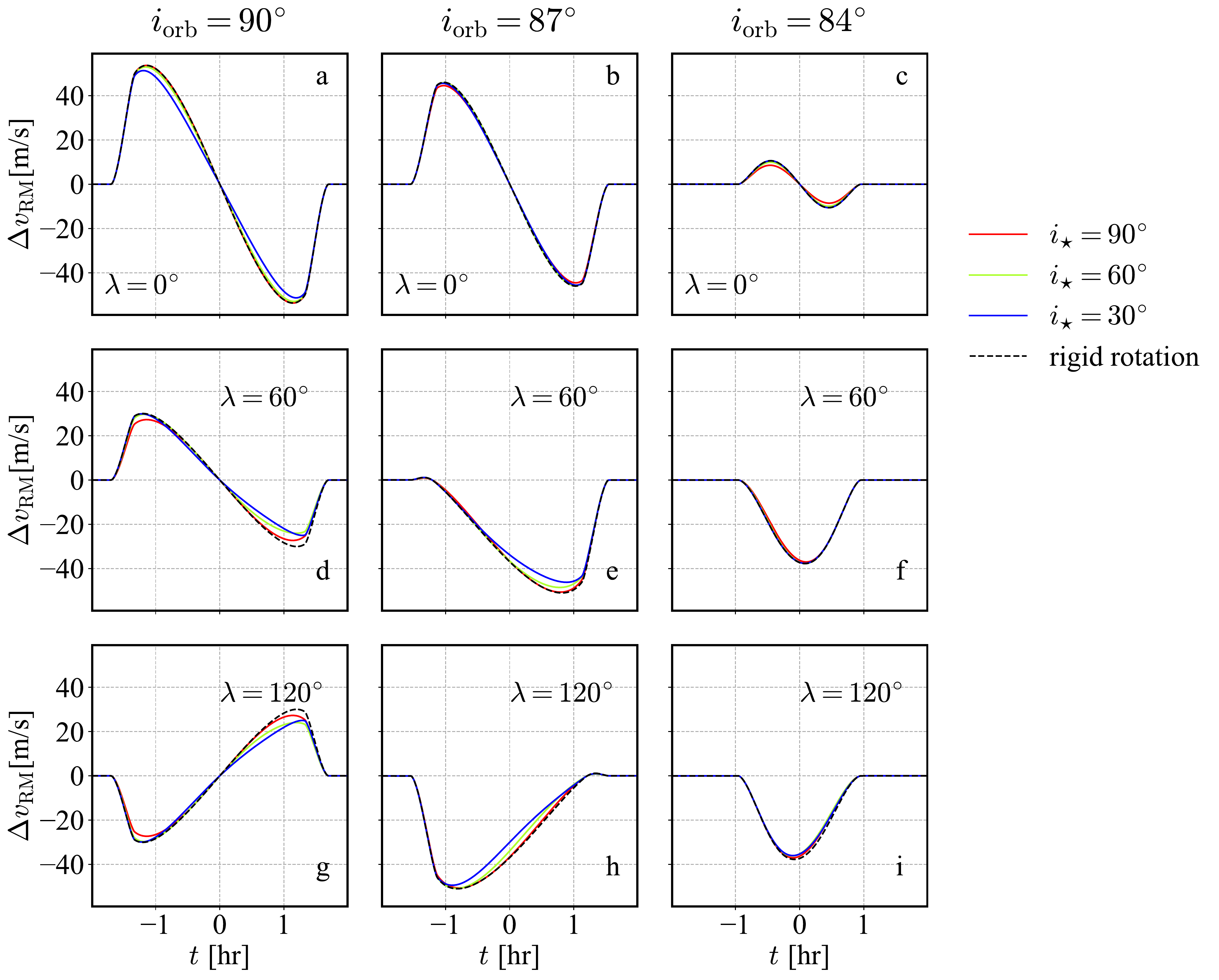} 
 \end{center}
\caption{Dependence of $\Delta v_{\rm RM}$ on the stellar inclination
  $\is$, where the equatorial velocity $\vsini$ is fixed ($4.5$km/s).
  Dashed lines correspond to the rigid rotation model. Left, middle,
  and right panels correspond to $\io=90^\circ (b=0)$, $87^\circ
  (b\approx 0.47)$, and $84^\circ (b\approx 0.93)$.  Red, greenyellow,
  and blue curves indicate $\is=90^\circ$, $60^\circ$, and $30^\circ$,
  respectively, for the differential rotation model with
  $\alpha_2=\alpha_{2\odot}$ and
  $\alpha_4=\alpha_{4\odot}$.} \label{fig:deltavrm}
\end{figure}

\subsection{Sensitivity to the stellar inclination}

\begin{figure}
 \begin{center}
  \includegraphics[width=12cm, bb = 0 0 1032 892]{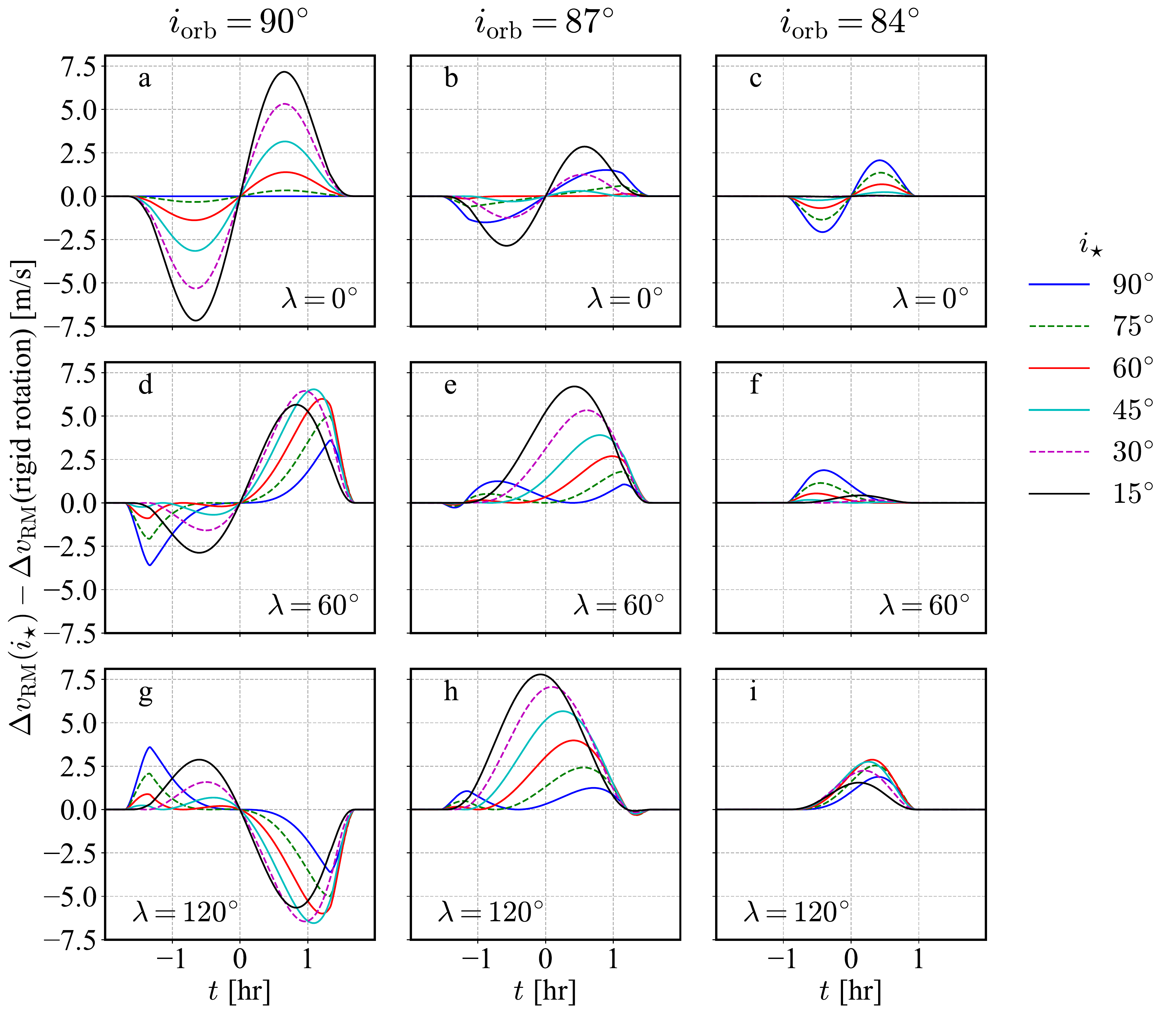} 
 \end{center}
 \caption{The RM modulation term due to the differential rotation
   after subtracting the rigid rotation component.  Each panel adopts
   the values of $\lambda$ and $\io$ in the corresponding panel of
   Figure \ref{fig:deltavrm}. We adopt $\alpha_2=\alpha_{2\odot}$
     and $\alpha_4=\alpha_{4\odot}$.  Blue-solid, green-dashed,
     red-solid, cyan-solid, purple-dashed, and black-solid curves
   correspond to $\is=90^\circ$, $75^\circ$, $60^\circ$, $45^\circ$,
   $30^\circ$, and $15^\circ$, respectively.
 }\label{fig:deltadeltavrm}
\end{figure}

Figure \ref{fig:deltavrm} compares $\Delta v_{\rm RM}$ with and
without differential rotation for different values of ($\io,
\lambda$).  Figure \ref{fig:deltadeltavrm} plots the corresponding RM
modulation term due to the differential rotation alone, $\Delta v_{\rm
  RM}(\is) - \Delta v_{\rm RM}({\rm rigid~rotation})$. In most cases,
the amplitude of the modulation term becomes the largest around the
ingress/egress phases, while it is suppressed by the limb darkening
depending on the specific values of $\io$ and $\lambda$.  The
modulation term vanishes at the central transit epoch if
$\io=90^\circ$ and/or $\lambda=0$ because $V_{\rm diff}(0) \propto
\cos\io \sin\lambda$ as is understood from equation
(\ref{eq:Vz-diff-0}).

\begin{figure}
 \begin{center}
  \includegraphics[width=9cm, bb = 0 0 1061 811]{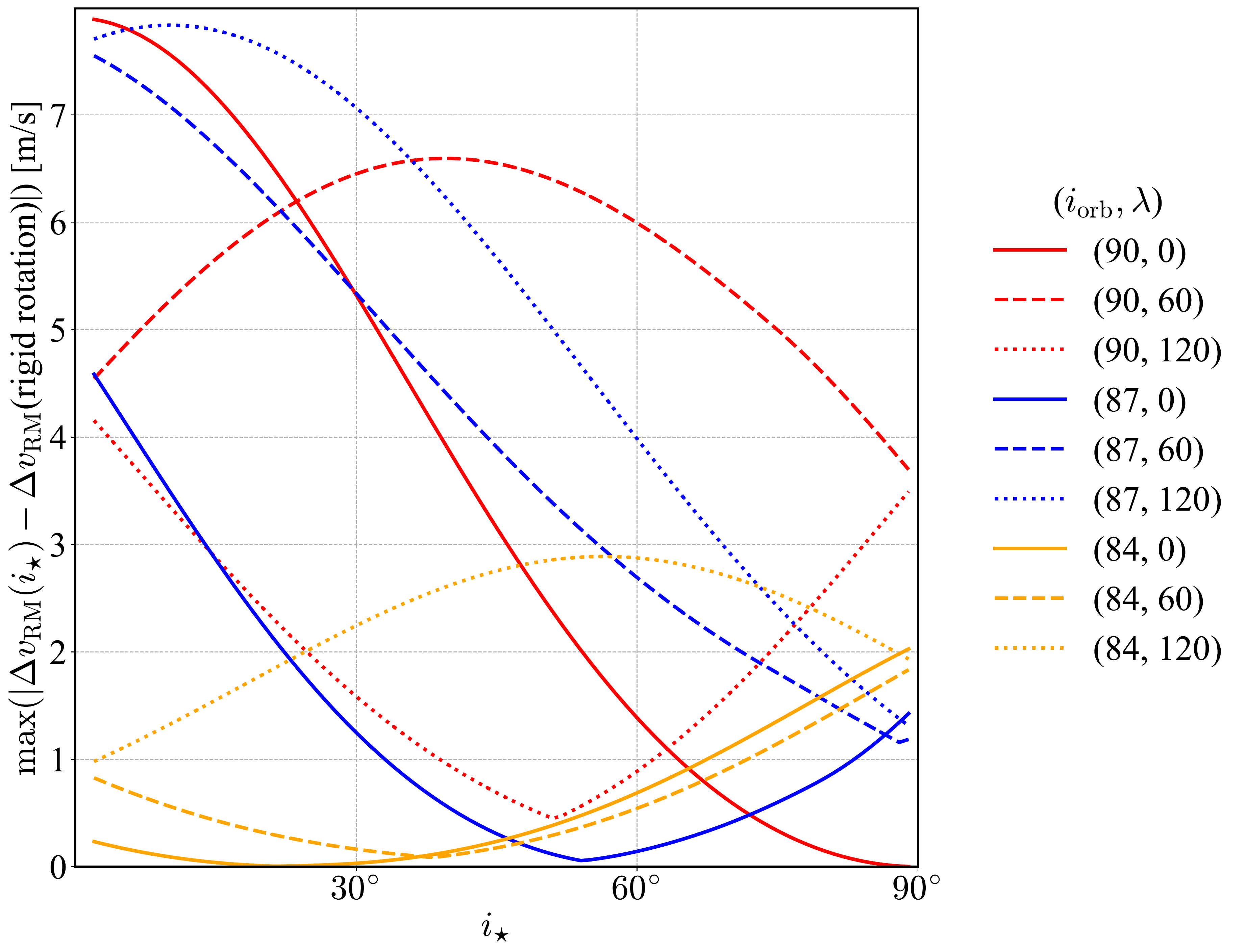} 
 \end{center}
 \caption{Maximum value of the RM modulation term due to the
   differential rotation during the entire transit period as a
   function of $\is$ for different values of $\io$ and $\lambda$
   indicated in the legend. We adopt $\alpha_2=\alpha_{2\odot}$
     and $\alpha_4=\alpha_{4\odot}$. Red, blue, and orange curves
     assume $\io=90^\circ (b=0)$, $87^\circ (b\approx 0.47)$, and
     $84^\circ (b\approx 0.93)$, while solid, dashed and dotted lines
     correspond to $\lambda=0^\circ$, $60^\circ$, and $120^\circ$,
     respectively.}
 \label{fig:diff-trend}
\end{figure}

Figure \ref{fig:diff-trend} plots the maximum values of the
RM modulation term during the entire transit period as a function of
$\is$.  If the degree of the differential rotation for the Sun is
assumed, the RM effect is supposed to have an additional modulation
component of an amplitude of several m/s relative to the rigid
rotation case. Given the current precision of the high-resolution
radial velocity measurement, it should be detectable for systems with
relatively high signal-to-noise ratios.

\subsection{Uncertainties of the spectroscopic parameters}

We have shown that, once the stellar differential rotation effect
  is taken into account, the precise measurement of the RM velocity
  anomaly $\Delta v_{\rm RM}$ can estimate both the stellar
  inclination $\is$ and the projected spin-orbit angle $\lambda$. So
  far, however, we assumed that the parameters $\beta$ and $\zeta$ in
  equations (\ref{eq:beta-*}) and (\ref{eq:beta-p}) are precisely
  determined from the stellar spectroscopy. In reality, however, the
  accurate determination of the two parameters is not so easy
  \citep[e.g.,][]{Kamiaka2018}.

\begin{figure}
 \begin{center}
  \includegraphics[width=12cm, bb = 0 0 1147 892]{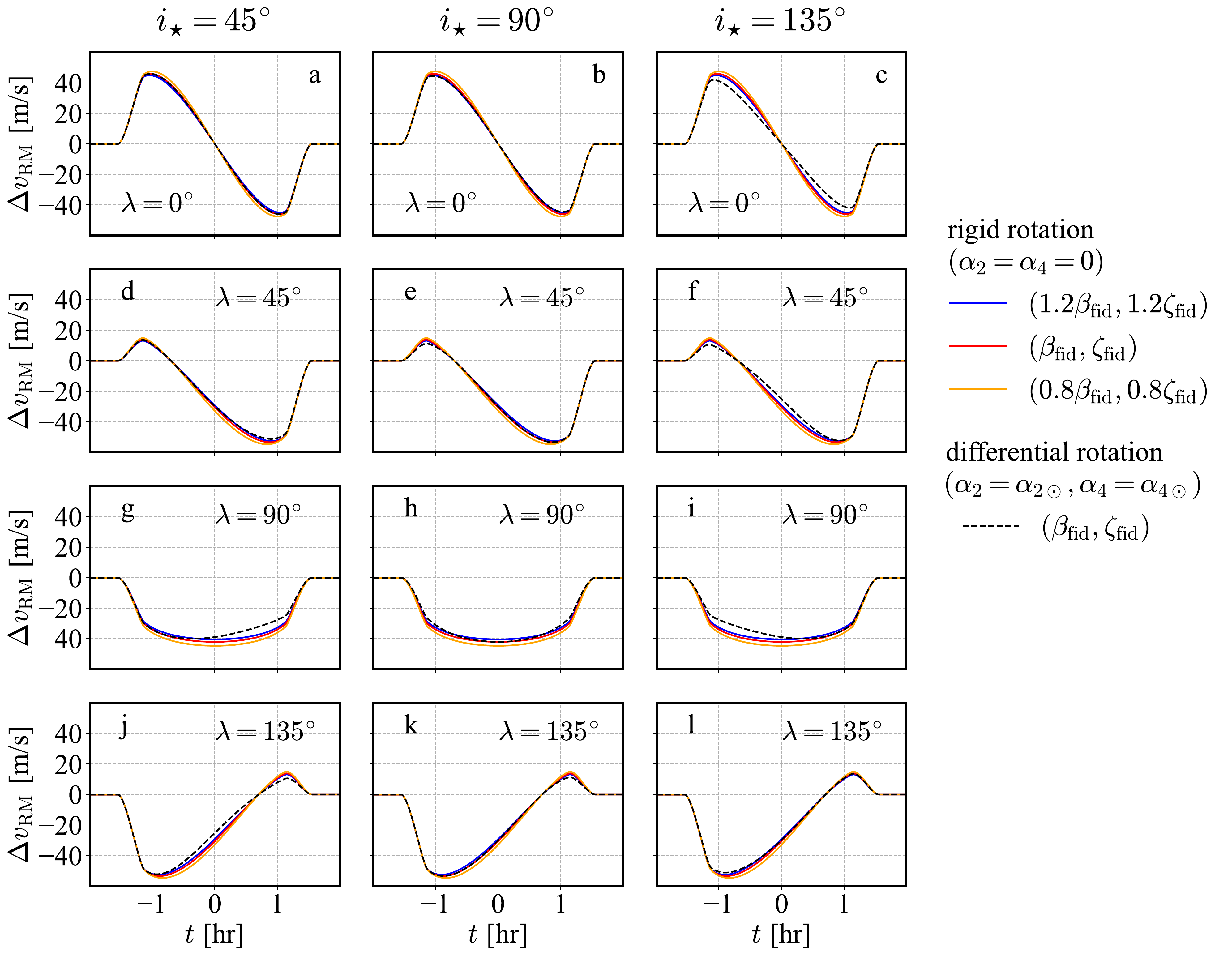} 
 \end{center}
 \caption{Same as Figure \ref{fig:I87} ($\io =87^\circ$), but for
     different sets of the values of $\beta$ and $\zeta$
     characterizing the spectroscopic line profiles.  Solid lines
     correspond to the rigid rotation model ($\alpha_2=\alpha_4=0$);
     blue, red, and orange curves correspond to
     $(\beta,\zeta)=(1.2\beta_{\rm fid}, 1.2\zeta_{\rm fid})$,
     $(\beta_{\rm fid}, \zeta_{\rm fid})$, and $(0.8\beta_{\rm fid},
     0.8\zeta_{\rm fid})$, respectively. For reference, the fiducial
     differential rotation model ($\alpha_2=\alpha_{2\odot}$,
     $\alpha_4=\alpha_{4\odot}$, $\beta=\beta_{\rm fid}$, and
     $\zeta=\zeta_{\rm fid}$) is plotted in dashed curves.  The
     fiducial values of the parameters are summarized in Table
     \ref{tab:paramvalue}.}
\label{fig:I87-betazeta}
\end{figure}

\begin{figure}
 \begin{center}
  \includegraphics[width=12cm, bb = 0 0 1149 890]{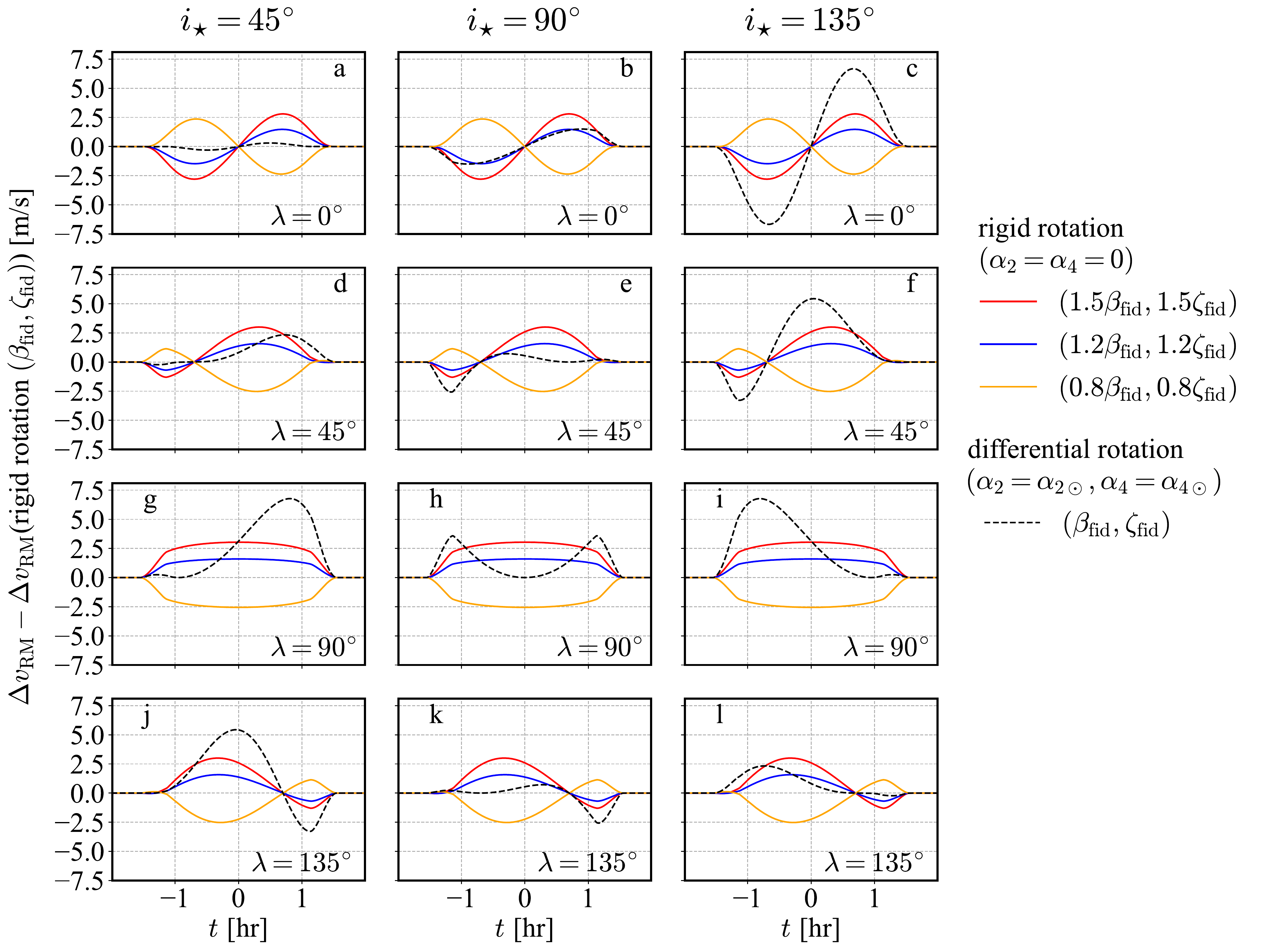} 
 \end{center}
 \caption{The RM velocity anomaly $\Delta v_{\rm
       RM}(\alpha_2,\alpha_4,\beta,\zeta)$ relative to that for the
     rigid rotation with fiducial spectroscopic parameters, $\Delta
     v_{\rm RM}(\alpha_2=\alpha_4=0, \beta_{\rm fid},\zeta_{\rm
       fid})$.  Solid lines correspond to the rigid rotation model
     ($\alpha_2=\alpha_4=0$); red, blue, and orange curves correspond
     to $(\beta,\zeta)=(1.5\beta_{\rm fid}, 1.5\zeta_{\rm fid})$,
     $(1.2\beta_{\rm fid}, 1.2\zeta_{\rm fid})$, and $(0.8\beta_{\rm fid},
     0.8\zeta_{\rm fid})$, respectively. For reference, the fiducial
     differential rotation model ($\alpha_2=\alpha_{2\odot}$,
     $\alpha_4=\alpha_{4\odot}$, $\beta=\beta_{\rm fid}$, and
     $\zeta=\zeta_{\rm fid}$) is plotted in dashed curves.}
\label{fig:I87-degeneracy}
\end{figure}

Thus it may be possible that the uncertainties of $\beta$ and
  $\zeta$ are degenerate with the stellar differential rotation
  effect. Since a comprehensive study of the parameter degeneracy is
  beyond the scope of this paper, we decided to show several examples.
  Figure \ref{fig:I87-betazeta} plots the dependence on $\beta$ and
  $\zeta$; strictly speaking, equations (\ref{eq:RM-Gaussian}),
  (\ref{eq:beta-p}), and (\ref{eq:beta-*}) indicate that $\Delta
  v_{\rm RM}$ depends on the combination of $\beta^2 + \zeta^2$ under
  the Gaussian approximation that we adopted throughout the
  paper. Comparison with Figure \ref{fig:I87} implies that while the
  uncertainties of the spectroscopic parameters distort the RM
  velocity anomaly curve, the distortion pattern is rather different
  from the signature due to the differential rotation.

For more quantitative comparison, we plot in Figure
  \ref{fig:I87-degeneracy} the residual of $\Delta v_{\rm
    RM}(\alpha_2,\alpha_4,\beta,\zeta)$ relative to that for the rigid
  rotation with fiducial spectroscopic parameters, {\it i.e.}, $\Delta
  v_{\rm RM}(\alpha_2=\alpha_4=0, \beta_{\rm fid},\zeta_{\rm fid})$.
  The resulting distortion due to the inaccuracy of the spectroscopic
  parameter is sensitive to the value of $\lambda$, but independent of
  the value of $\is$, while the signature of the differential
  rotation model is sensitive to both $\lambda$ and $\is$.  Figure
  \ref{fig:I87-degeneracy} implies that if the RM velocity anomaly is
  determined with a precision on the order of m/s, one can identify
  the differential rotation signal even under the presence of a
  relatively large uncertainty of $\beta$ and $\zeta$.

\section{Summary and conclusions}

The Rossiter-McLaughlin (RM) effect has unveiled an unexpected
diversity of the spin-orbit architecture of transiting planetary
systems. Most of the previous approaches have assumed a rigid rotation
of the host star, and thus were not able to estimate the stellar
inclination $\is$ except for the combination of $\vsini$.  We have
improved the previous model of the RM effect by incorporating the
differential rotation of the host star. Since our model is fully
analytic, it provides a good theoretical understanding of how one can
determine $\lambda$ and $\is$ simultaneously from the differential
rotation effect.

The latitudinal differential rotation of planetary host stars
  leaves an additional characteristic signature on the RM effect
  through equation (\ref{eq:Vz-diff}). More specifically, under the
  Gaussian approximation of stellar lines, the RM effect is
  analytically described by equation (\ref{eq:RM-Gaussian}), together
  with equations (\ref{eq:beta-p}), (\ref{eq:beta-*}),
  (\ref{eq:f-formula}), and (\ref{eq:vp-formula}). We can estimate the
  values of the spectroscopic parameters, $\beta_{\rm p}$ and
  $\beta_\star$, and the limb darkening parameters, $q_1$ and $q_2$
  from photometric and spectroscopic observations of the host star.
  If the differential rotation of the star is absent, equation
  (\ref{eq:Vz-diff}) vanishes, and equation (\ref{eq:Vz-rigid}) is
  specified by $\vsini$ and $\lambda$. This provides a widely used
  method to estimate the projected spin-orbit angle $\lambda$ from the
  RM effect. The differential rotation term, equation
  (\ref{eq:Vz-diff}), makes it possible to estimate $\alpha_2$,
  $\alpha_4$ and $\is$ separately.

The amplitude of the RM modulation term due to the differential
rotation is on the order of several m/s for stars similar to the Sun,
and thus should be detectable for systems with reasonably high
signal-to-noise ratio of the RM measurement.  Therefore, our model is
useful in recovering the full spin-orbit angle $\psi$, equation
  (\ref{eq:cospsi}), based on the RM data analysis alone in
  principle. In reality, the degeneracy among several parameters might
  complicate the interpretation of the differential rotation feature
  in a robust fashion. A preliminary result shown in Figure
  \ref{fig:I87-degeneracy} implies, however, that the uncertainties
  due of the spectroscopic line profiles are distinguishable from the
  real differential rotation signature, depending on the values of
  $\is$ and $\lambda$ of specific systems.  We plan to perform a
  further study of the parameter degeneracy, and to apply the current
  methodology to a sample of transiting planetary systems, which will
  be discussed elsewhere.

\bigskip

\section*{Acknowledgements}
We thank an anonymous referee and Teruyuki Hirano for many useful
  and constructive comments.  The present work is supported by
Grants-in Aid for Scientific Research by the Japan Society for
Promotion of Science (JSPS) No.18H012 and No.19H01947, and from JSPS
Core-to-core Program ``International Network of Planetary Sciences''.

\bibliographystyle{apj}
\bibliography{ref-suto, ref-sasaki} 

\end{document}